\documentclass[sigconf]{acmart}

\AtBeginDocument{%
  \providecommand\BibTeX{{%
    \normalfont B\kern-0.5em{\scshape i\kern-0.25em b}\kern-0.8em\TeX}}}

\usepackage{booktabs}
\usepackage{pbox}
\usepackage{epsfig,endnotes}
\usepackage{epstopdf}
\usepackage{xcolor}
\usepackage{graphicx}
\usepackage{xspace}
\usepackage{enumitem}
\usepackage{url}
\usepackage{balance}
\usepackage{caption}
\usepackage{subcaption}
\usepackage{breakurl}
\usepackage{pbox}
\usepackage{times}  
\usepackage{pdfrender}
\usepackage{csquotes}
\usepackage{amsmath}

\setcopyright{none}
\renewcommand\footnotetextcopyrightpermission[1]{}
\setcopyright{none}
\makeatletter
\renewcommand\@formatdoi[1]{\ignorespaces}
\makeatother

\begin{document}

\newcommand{\argmax}[1]{\underset{#1}{\operatorname{arg}\,\operatorname{max}}\;}

\newcommand{\eg}{e.g.,\xspace}
\newcommand{\etc}{etc.\xspace}
\newcommand{\etal}{et al.\xspace}
\newcommand{\ie}{i.e.,\xspace}
\newcommand{\point}[1]{\vspace{.05in}\par\noindent\textbf{#1}:\xspace}

\title{Differential Tracking Across Topical Webpages of Indian News Media}

\newcommand\sbm[1]{\textbf{\textcolor{brown}{SBM: #1}}	}
\newcommand\nk[1]{\textbf{\textcolor{blue}{NK: #1}}	}
\newcommand\ns[1]{\textbf{\textcolor{red}{NS: #1}}	}
\newcommand\pa[1]{\textbf{\textcolor{purple}{PA: #1}}	}
\newcommand\yv[1]{\textbf{\textcolor{orange}{YV: #1}}	}
\newcommand\va[1]{\textbf{\textcolor{green}{VA: #1}}	}

\author{
Yash Vekaria\texorpdfstring{$^\star {} ^\alpha$}{},
Vibhor Agarwal\texorpdfstring{$^{\star} {} ^{\alpha}$}{},
Pushkal Agarwal\texorpdfstring{$^\beta$}{},
Sangeeta Mahapatra\texorpdfstring{$^\gamma$}{},\texorpdfstring{\\}{}
Sakthi Balan Muthiah\texorpdfstring{$^\alpha$}{},
Nishanth Sastry\texorpdfstring{$^\delta$}{},
Nicolas Kourtellis\texorpdfstring{$^\zeta$}{}
}

\affiliation{
$^\alpha$The LNM Institute of Information Technology, Jaipur 
\country{India}\\
$^\beta$King's College London, London 
\country{United Kingdom}\\
$^\gamma$German Institute for Global and Area Studies, Hamburg 
\country{Germany}\\
$^\delta$University of Surrey, Surrey
\country{United Kingdom}\\
$^\zeta$Telefonica Research, Barcelona \country{Spain}\\
$^\alpha$\{yash.vekaria.y16, vibhor.agarwal.y16,  sakthi.balan\}@lnmiit.ac.in,
$^\beta$pushkal.agarwal@kcl.ac.uk\\
$^\gamma$sangeeta.mahapatra@giga-hamburg.de,
$^\delta$n.sastry@surrey.ac.uk,
$^\zeta$nicolas.kourtellis@telefonica.com}

\begin{abstract}

Online user privacy and tracking have been extensively studied in recent years, especially due to privacy and personal data-related legislations in the EU and the USA, such as the General Data Protection Regulation, ePrivacy Regulation, and California Consumer Privacy Act.
Research has revealed novel tracking and personal identifiable information leakage methods that first- and third-parties employ on websites around the world, as well as the intensity of tracking performed on such websites.  
However, for the sake of scaling to cover a large portion of the Web, most past studies focused on homepages of websites, and did not look deeper into the tracking practices on their topical subpages.
The majority of studies focused on the Global North markets such as the EU and the USA.
Large markets such as India, which covers 20\% of the world population and has no explicit privacy laws, have not been studied in this regard.

We aim to address these gaps and focus on the following research questions: Is tracking on topical subpages of Indian news websites different from their homepage?
Do third-party trackers prefer to track specific topics?
How does this preference compare to the similarity of content shown on these topical subpages?
To answer these questions, we propose a novel method for automatic extraction and categorization of Indian news topical subpages based on the details in their URLs.
We study the identified topical subpages and compare them with their homepages with respect to the intensity of cookie injection and third-party embeddedness and type.
We find differential user tracking among subpages, and between subpages and homepages.
We also find a preferential attachment of third-party trackers to specific topics.
Also, embedded third-parties tend to track specific subpages simultaneously, revealing possible user profiling in action.

\end{abstract}

\maketitle

\renewcommand{\thefootnote}{}
\footnote{\texorpdfstring{$^\star$}{}Both the authors contributed equally to this research.}
\renewcommand{\thefootnote}{\arabic{footnote}}
\lhead{Differential Tracking Across Topical Webpages of Indian News Media}
\rhead{Vekaria Y. and Agarwal V., et al.}

\section{Introduction}
India has seen exponential growth in online users with the recent advancements in network coverage and cheap internet rates.
This enormous growth coupled with the lack of a data privacy law in India\footnote{Personal Data Protection Bill, tabled in Indian Parliament in December 2019, is still with the Joint Parliament Committee for review.} potentially allows for unregulated tracking of users. 
Though tracking has its
advantages, it also gives opportunities to the third-parties to
track the readers and build their profiles, which can be used further
for targeted advertisements and influence campaigns.
Studies have shown that news websites have more tracking than other categories~\cite{englehardt2016census,agarwal2020stop}.
Therefore, we choose to focus on Indian news websites to demonstrate tracking in this under-studied country, which happens to be the second largest internet market in the world~\cite{internet-usage}.

An important observation of recent studies~\cite{aqeel2020landing,urban2020beyond} is that researchers have been focusing on the homepages of websites for understanding the tracking ecosystem around the world due to its ease in scaling the approach.
Our focus on one country (India) and one type of websites (news), allows us to dig deeper to understand how tracking varies across topical subpages of a website.
We believe this is an orthogonal, yet important contribution that needs to be replicated in other, better studied settings such as the USA and the European Union.
News is ideally suited for such a study across different subpages of a website since different users are likely to be interested in different topics.

To this end, we undertake, to our knowledge, the first in-depth study of Web tracking in more than 100 top Indian news websites\footnote{Our dataset and code is released via this link for non-commercial research usage: \url{http://tiny.cc/india-topic}.}. 
In this work, we address three research questions: (1) Is there any difference in tracking on topical subpages of Indian news websites and their homepages? (2) Do third-party trackers have preferential attachment to specific topical subpages? (3) Does this preference have any connection with the similarity of content shown to users on topical subpages?
Towards addressing these questions, we propose DiBETS, a method for classifying topical categories based on keywords of their URLs, and demonstrate differences in tracking of 112 homepages of websites and across 1.3K topical subpages.

We find that 11 out of 15 topical subpages are \textit{tracked more intensely} than their homepages -- topical subpages can have as high as 44\% more cookies.
For some topical subpages, there are additional third-parties (TP), which are involved in more fingerprinting and analytics when compared to their homepages.
We also find that there is a \textit{specialisation} of third-party partners, who prefer to advertise or track in specific topics.
Some of these appear to be aligned to specific business affinities of the trackers for different topics.
For example, music websites, which track or advertise on the `entertainment' subpage, or games and sports websites advertising on `sports' subpage.
We also find some counter-intuitive preferences, such as a paint company tracking in Coronavirus-related subpages.
We find that such aberrations can be explained in some cases (in this case, the paint company recently claimed to have released an {\em anti-Covid-19 paint}\footnote{https://realtyplusmag.com/asian-paints-launches-anti-covid-19-paint/} and also introduced sanitisers\footnote{https://english.jagran.com/business/exclusive-asian-paints-fights-covid-crisis-with-safe-painting-and-san-assure-services-focuses-on-health-and-hygiene-products-10014656} in its Health and Hygiene segment during the start of the pandemic).
Using clustering approaches, we show that embedded TP trackers tend to track specific subpages simultaneously, revealing possible user profiling in action.
Such preferential attachments may be among smaller players who have affinities towards specific topics and subpages.
However, major TP players (such as {\tt doubleclick.net}) are  present on all topics. 
Finally, we note that privacy policies are nearly always in English, even on regional Indian language news websites, which make it hard for the audience to truly understand or consent to ad-related privacy policies.

\section{Related Work and Background}
\label{sec:related_works}

Previous studies focused on tracking in different categories of websites (e.g.,~News~\cite{urban2020beyond, englehardt2016census}, Adult~\cite{vallina2019porn-imc}, Education, etc.) and types of users.
In~\cite{agarwal2020stop} authors showed that tracking on various demographics, simulated by artificial user personas using categorical websites list from Alexa Inc.~\cite{Alexa}, can significantly differ.
For instance, after visiting around 20 websites related to youth and women, there were about 60 and 300 distinct TP trackers, respectively, in an artificial user's history.
Among all personas, youth demographics had the least number of trackers, possibly because of strict laws such as USA's Children Online Privacy Protection Rule~\cite{COPPA}, which protects the youth population from being tracked.
In~\cite{urban2020beyond} authors used McAfee SmartFilter service~\cite{McAfcc} to obtain categories of one million top trending websites.
They found that tracking on news websites was more than other types~\cite{urban2020beyond, englehardt2016census}.

Interestingly, internal page discovery and analysis have only been recently studied~\cite{aqeel2020landing,urban2020beyond}.
Subpages of a website cover various different topics, and can be based on user preferences including trends and themes of the website~\cite{aqeel2020landing}, as well as catering to a user's interest in `hard' or `soft news' with their differing content and style of presentation~\cite{Reinmann2012HardSoft}.
These works emphasize the need for deeper analysis of internal webpages, which differ in content, performance, and user tracking from the landing/home page.
In~\cite{aqeel2020landing}, the landing page of a website was compared with several internal pages of the same website in order to analyse the difference in the content and the performance from the landing page. 
This was done for 20K pages from about 1K websites.
Their main contribution was to show that the Web performance measurement and optimization need not be only based on the landing pages but also on the internal pages. 
In our work, the main objective is to look at the tracking ecosystem in the topical subpages of news websites.
Our focus is more related to~\cite{urban2020beyond}, where they analyse the relationship between third-parties in over 10K websites.
They build third-party trees that capture the loading sequences and the dependencies of various third-parties.
Their study indicates that there are substantially more cookies in the subpages than the site’s landing page (36\% more), which implies that studies based only on the landing pages could be biased.

There are various studies which use different computational methods to quantify and understand topics on subpages, such as Word2Vec in the Indian context for Tamil language~\cite{ramraj2020topic}, Na\"ive Bayes in Arabic language~\cite{fouad2018efficient}, etc.
These topic identification techniques can help publishers in providing personalised news and understanding the emergence of new topics over time~\cite{xu2019research, katyayani2019hot,10.1145/1242572.1242610}.
In fact, previous studies have focused on personalised news topic selection based on user profiles and browsing history, for e.g., using click behaviour~\cite{Liuetal2010Personalised}, and public news selection through raw text analysis~\cite{Toraman2015Frontpage}.
The present study is different as it is not about topic selection or salience from end user perspective but one derived from the presence of Web trackers on topical subpages. We study
which topics attract most trackers for user targeting.  
Furthermore, news topic analyses are usually related to media slant, i.e.,~how news publications tilt towards ideological positions based on readers preference, ownership consideration, and market compulsions~\cite{gentzkow2010drives}, or agenda setting function of news media, i.e.,~how mass media influences public opinion on certain issues/agendas~\cite{Rogers1988Agenda}.
Our study on tracking of topical subpages can aid these studies as trackers can provide detailed information on the profiles of the users and their associated browsing behaviour, allowing for targeted ads and slanted news coverage targeted at users.

Indeed, while topical tracking has been studied across websites, tracking on topical subpages within websites is still fairly unexplored.
Hence, we build upon past studies, to present topical tracking in Indian news websites.
We focus on Indian news as it has huge amount of trackers compared to the general Web.
A reason can be that India witnessed massive growth in Internet users in the past few years~\cite{agarwal2021under}.
In general, such Indian news websites have different categories of news in their homepage, such as International, National, Politics, Business, Sports, etc.
The collection of data from homepages and their subpages in automated way is prone to encounter problems since the topical names might be different in various websites.
For example, in the Times of India, there is a category called {\it India}, but in The Hindu, the same semantically-related category is called {\it National}.
Moreover, there are new trending topics in most Indian newspapers like {\it Coronavirus/Covid}, and there are examples of topics such as {\it Manoranjan}, which is a Hindi word for 'entertainment'.
So, in order to collect the data and segregate it into categories, we built a new model called {\it DiBETS} (Dictionary-Based Extractor of Topical Subpages) which classifies the subpage URLs into different categories. This supplements more manual and qualitative approaches previously undertaken in the field~\cite{newman2015reuters}. An additional differentiation from such previous works is that they do not focus on the Indian News landscape, which is very large and highly diverse.

\section{Methodology and Dataset Collection}

In this section, we describe our methodology to study tracking in topical subpages.
We first provide details on how we identify topical subpages per website using DiBETS (Dictionary-Based Extractor of Topical Subpages), our novel method (Sec.~\ref{sec:methodology-dibets}).
Next, we discuss application of DiBETS to Indian news context (Sec.~\ref{sec:methodology-dibetsapplication}) and outline our tracking and content crawling of webpages (Sec.~\ref{sec:methodology-crawling}).

\begin{figure*}
\centering
  \includegraphics[width=\textwidth]{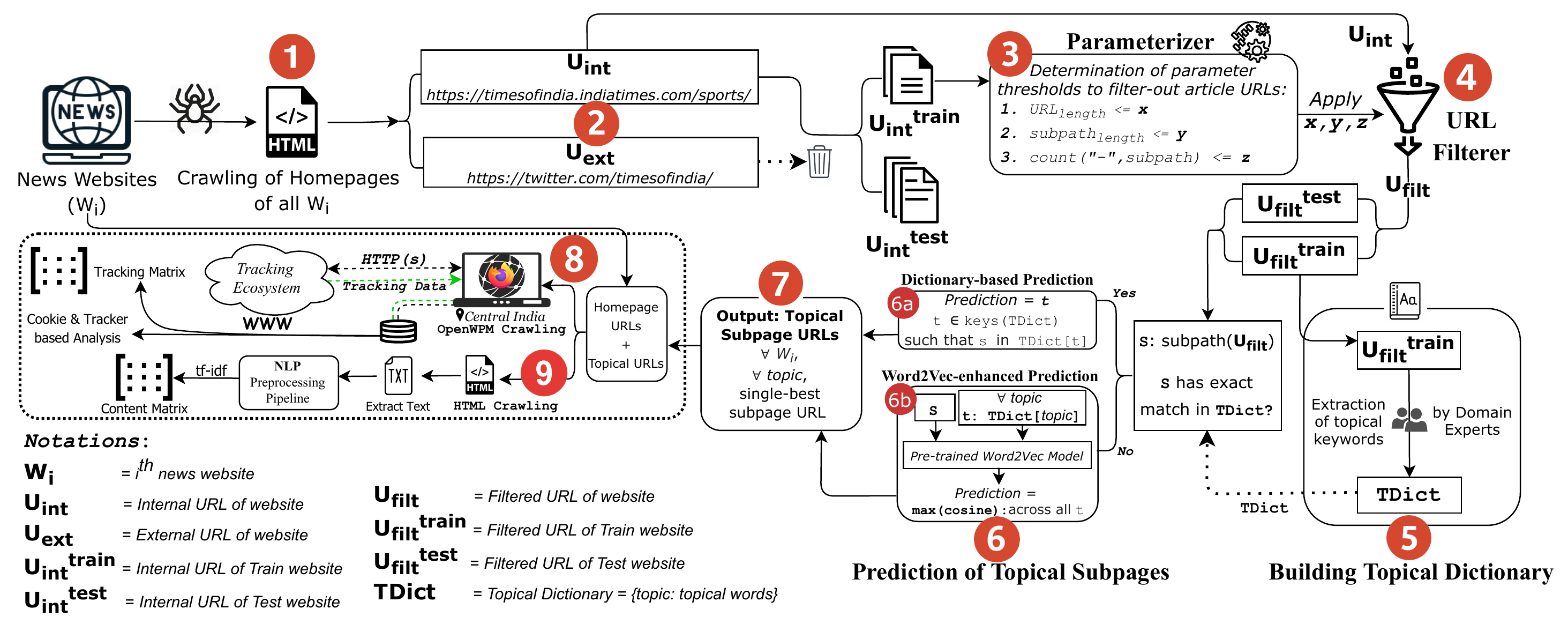}
  \caption{Our methodology to study tracking of topical subpages in news websites.}
  \label{fig:DiBETS}
\end{figure*}

\subsection{Automated Extraction of Topical Subpages with DiBETS}
\label{sec:methodology-dibets}

We propose DiBETS, a Word2Vec-enhanced method to extract the main topical subpages from a website's homepage.
Since our study focuses on news websites, we tune DiBETS to detect the subpages of 15 standard topics covered on news websites.
We choose these topics from a study by Reuters~\cite{newman2015reuters}, where it compares the focus of 12 topics in the news websites of eleven countries.
We regroup some topics used by~\cite{newman2015reuters} taking a higher level perspective (e.g., \textit{Business and Finance} and \textit{Economics} those being related, are grouped into one;  \textit{Fun/Weird News} is included under \textit{Entertainment}). 
We also add few other categories: e.g.,  \textit{Multimedia} is more usefully seen as a separate topic based on the prominent presence of multimedia-based subpages  on Indian news websites.  
In the context of Covid-19, numerous news websites are seen to incorporate a separate subpage for \textit{Coronavirus}; hence, we define it as a separate topic.
To study the privacy pages and cookie policy in Indian context, we also add a topic of \textit{Privacy Policy}.
Thus, we obtain a master list of 15 topics. 
We classify a subpage that belongs to topics outside these 15 topics as \textit{Other}. 
Figure~\ref{fig:DiBETS} illustrates our methodology, including DiBETS' main Steps 1-7, and the subsequent data crawling Steps 8-9.

\subsubsection{\textbf{Website Crawling and Extraction of Internal URLs}}

DiBETS takes a news website ($W_{i}$) as an input and first crawls the raw HTML of the website's homepage (\textit{Step 1}).
In \textit{step 2}, from the raw HTML of the homepage, all valid embedded URLs are extracted (HTML ``href'' tags).
These URLs are segregated into internal and external URLs.
Internal URLs are the ones belonging to the same domain as the homepage.
External URLs point to a different domain and are discarded.
These two steps are performed collectively for all websites of interest.
We select a small portion of websites (and its URLs) to train and tune DiBETS in a supervised fashion, and then apply it on the remaining websites.

\subsubsection{\textbf{URL Parameterizer and Filterer}}
\label{sec:methodology-parameterizer}
Internal URLs may consist of article URLs or other subpage URLs.
Since our objective is to obtain the subpage URLs of different topics focused on the homepage of a website, we filter out article URLs.
Manual inspection of a subset of the training set revealed that article URLs are typically longer with respect to number of characters and subpaths (subpath is a portion of URL path between two consecutive ``/'', each organized as hyphen-separated words) and have more hyphens than subpage URLs.
In fact, these three parameters (\emph{URL length}, \emph{subpath length} of subpaths within that URL, and \emph{hyphen count} within a subpath) have clear bi-modal distribution, easily separating the two types of URLs (article vs. subpage URLs).
Thus, we identify automatically a threshold for each parameter (i.e., \textit{x}, \textit{y}, and \textit{z} respectively for $\emph{length(\texttt{URL})}$, $\emph{length(\texttt{subpath})}$ and $\emph{count(``-'', \texttt{subpath})}$), as the point between the two modes in the frequency distribution of each parameter.
\textit{Step 4} applies these thresholds and keeps as subpage URLs the internal URLs that have, for all three parameters, numbers that are less than or equal to the thresholds found.

\subsubsection{\textbf{Topical Dictionary}}
In \textit{Step 5}, we build the Topical Dictionary by adapting one-time manual process with the help of a domain expert who is well-versed with the thematic structure followed by news websites.
The dictionary is built with the remaining subpage URLs of the train set from \textit{Step 4}.
For each URL, the expert attributes the URL to one of the 16 (15 topical subpages plus `Other') by either looking at the URL or actually visiting the URL and observing its content.
Additionally, the expert associates keywords present in the URL's subpaths
that can be suggestive of the subpage's topic.
Following this, we generate a dictionary (\texttt{TDict}) consisting of (\texttt{key}, \texttt{value}) pairs, where key is any of the 16 topics, and value is a list of keywords found in the subpage URLs of the said topic.

\subsubsection{\textbf{Prediction of Topical Subpages}}
To determine the topic of a URL in \textit{Step 6}, the subpaths of each filtered URL are considered hierarchically from high-level subpath to a low-level, one by one, and checked in the Topical Dictionary.
If the subpath is a generic phrase (e.g.,``category'' or ``section''), we continue to the next one.\\
\noindent \textbf{Dictionary-based Prediction (Step 6a):} If there is an exact match of the subpath with any keyword in the Topical Dictionary, then the topic (key) of that keyword is assigned to that subpage URL.
Else, we use \textit{Step 6b} to get a topic.\\
\noindent \textbf{Word2Vec-enhanced Prediction (Step 6b):} DiBETS uses a pre-trained Word2Vec model on English CoNLL17 corpus with a vocabulary size of 4027169
words~\cite{kutuzov2017word}.
This model is trained using a Word2Vec Continuous Skip-gram algorithm with a vector size 100 and a window of 10 words.

To predict the topic of a subpage URL, first, we tokenize the subpath, remove stop words, and compute a combined word embedding $comb\_emb[subpath]$ for all tokens found.
Similarly, for each topic \texttt{$T_i$}, we also pre-compute a combined word embedding $comb\_emb[T_i]$ by summing up word embedding of all its keywords in \texttt{TDict}.
Finally, we compute the cosine similarity between these two embeddings. The predicted topic is determined as per Eq~\eqref{eqn:predicting_topic}, if the corresponding maximum value of cosine similarity is greater than the threshold value.
The same automated method from Sec.~\ref{sec:methodology-parameterizer} is used to set the threshold between the bi-modal histogram of maximum cosine score amongst the cosine scores of each internal URL with all the topics.
\begin{equation*} 
\label{eqn:predicting_topic}
    T_{pred} = \underset{T_i \in keys(TDict)}{\arg\max} cosine(comb\_emb[subpath],\ comb\_emb[T_i]) \tag{ii}
\end{equation*}

\subsubsection{\textbf{Determining Single Best subpage URL for each topic}}
Each filtered URL is assigned one of the 16 topics by DiBETS.
As a result, there may exist more than one subpage URL (of the same website) for a given topic in the output of \textit{Step 6}\footnote{E.g. \url{https://www.xyz.com/sports/}, \url{https://www.xyz.com/sports/cricket/} and \url{https://www.xyz.com/sports/athletics/} are all assigned to topic ``Sports''.}.
So in the final stage (\textit{step 7}), the model determines and outputs a single best subpage URL for each topic per website (depending upon the topics that are detected by the model in that website).
Out of all the URLs assigned to the same topic \texttt{T}, DiBETS weighs each URL by the following ratio\footnote{Shorter the URL, higher the probability of it being the main subpage URL.
This is because shorter URLs have single or smaller hierarchy of subpaths and represents the main subpage for a topic.
Longer URLs add more hierarchy to subpaths, resulting in sub-subpage URLs.
Hence, the ratio is inversely weighted by the number of tokens in subpath after stop-word removal.}:
\begin{equation*} 
\label{eqn:ranking_weight}
    \textit{weight} = \frac{cosine(comb\_emb[all\ subpaths],\ comb\_emb[TDict[T]])}{count(tokens\ across\ all\ subpaths)} \tag{iii}
\end{equation*}

These URLs are then ordered in descending order of the weight Eq~\eqref{eqn:ranking_weight} and checked from the top if the URL's highest-level subpath is entirely present in our Topical Dictionary (corresponding to the assigned topic \texttt{T}) or not. If yes, DiBETS outputs it as the main subpage URL of \texttt{T} for a website.
Otherwise, it returns the first URL in the above ordering.

\subsection{\textbf{DiBETS application to Indian News Context}}
\label{sec:methodology-dibetsapplication}

The main input to our DiBETS model is the raw HTML of website homepages.
We used a dataset we collected previously, of 123 popular Indian news websites~\cite{agarwal2021under}.
Using an automated Python script, we performed two separate crawls of the homepages of all 123 websites in July and December 2020.
We compared the output of DiBETS across both crawls and the results of topical subpages were stable across time and topics (fraction of websites of topic $i$ in July and December remained almost constant, with a difference of $\sim4\%$ (min=0.98\%, max=22\%)).
Our analysis in Sec.~\ref{sec:topical_tracking} is on the December 2020 crawl.

\begin{figure*}
\centering
  \includegraphics[scale=0.5]{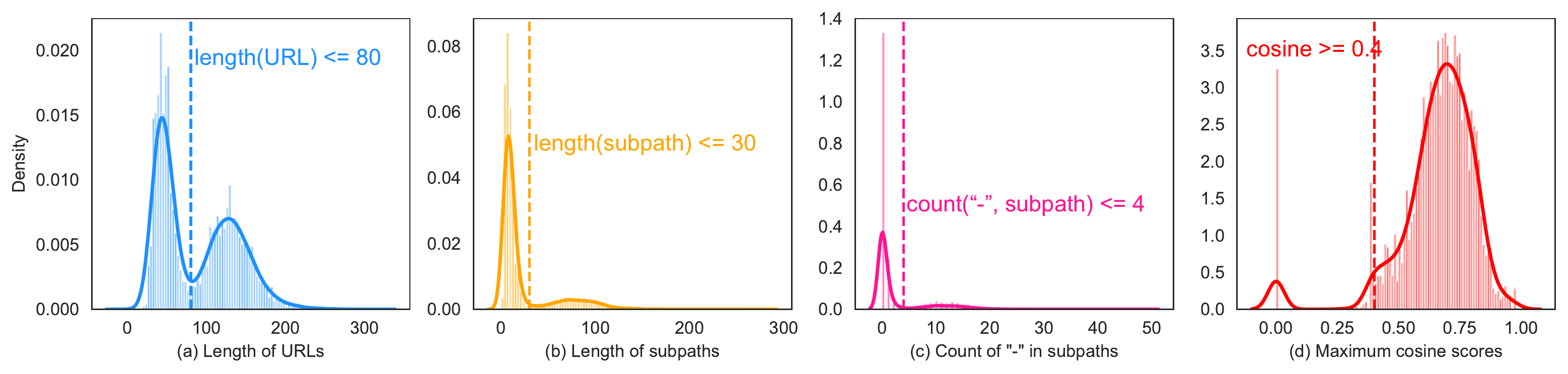}
  \caption{Histogram distributions of URL-specific parameters: (a) Length(URL), (b) Length(subpath), (c) Count(`-', subpath), (d) Histogram of Word2Vec-based maximum cosine similarity scores of all the internal URLs. Finalized thresholds for each distribution are marked by vertical dashed lines.
    }
  \label{fig:histogram}
\end{figure*}

After passing the 123 websites to \textit{Step 1} of DiBETS, we could retrieve the HTML of only 112 websites (11 websites forbade access).
In them, we found 15540 internal and 2718 external URLs.
Then, we identified the following as thresholds for the three parameters involved:
$\emph{length(\texttt{URL})} \leq \textit{80}$,
$\emph{length(\texttt{subpath})} \leq \textit{30}$, and
$\emph{count(``-'', \texttt{subpath})} \leq \textit{4}$.
Also, the maximum cosine score threshold in the Word2Vec (Step 6b) was found to be 0.4 (also shown in Figure~\ref{fig:histogram}).
The Topical dictionary built comprises a total of 1491 distinct keywords. Thus, as the final output of DiBETS in \textit{Step 7}, we obtained 1275 subpage URLs for different topics detected across sites.

\noindent \textbf{Evaluation of DiBETS Performance.}
\label{sec:methodology-generalisability}
We asked two domain experts from political science and journalism backgrounds to verify the prediction of the best topical subpages across 112 websites by DiBETS.
Each expert was presented with the model's output of a randomly selected set of 250 subpage URLs alongside the topic predicted by the model.
Then, they were independently asked to make a binary decision whether the model selection for best topical subpage was correct or not.
We computed 96\% agreement between experts, and an inter-annotator agreement with Cohen's Kappa score of 0.66, representing a ``substantial agreement''.

\noindent \textbf{Generalisability of DiBETS.}
To test the generalisability of our model, we also applied DiBETS in the USA context on a set of top 15 American news websites (based on Alexa ranks and Facebook followers), crawled in January 2021.
We applied similar process for training and testing the method and asked our experts to perform annotation for these results as well.
We observed an agreement of 95.67\% and Cohen's Kappa as 0.72.
These results give us a confidence that DiBETS is applicable in other contexts outside India.

\subsection{\textbf{Website Crawling for Tracking and Content}}
\label{sec:methodology-crawling}

As depicted in Step 8 of Figure~\ref{fig:DiBETS}, we crawled websites and collected traffic data using OpenWPM~\cite{englehardt2016census, openwpm-code}, a web privacy measurement framework.
We used most settings at default, with Firefox browser and third-party tracking enabled and tracking protection as false so that nothing is blocked.
We also enabled http\_instrument which logs HTTP(s) responses, requests and redirects, using a selenium headless browser to perform crawling and setting javascript\_instrument, and javascript\_cookies to true to store javascript cookies information.
We performed five stateless crawls of the 112+1275=1387 URLs (output from DiBETS), consisting of news websites' homepages and topical subpages during 1-31 December 2020. In total, we logged more than 1.2M cookies, 1.7M HTTP(s) requests, 1.5M responses, and 297K redirects.
Since stateless crawls make each website visit independent, parallel browser instances were launched to allow simultaneous crawls of these news websites from Central India. 
Moreover, we performed such crawls with a gap of six days to log the dynamic traffic data and account for infrequent but unavoidable network errors during each crawl.

Using the above data, we mapped the presence of 561 distinct third-party trackers found across homepages and topical subpages as a 16x561 binary matrix, which we call \textit{tracking matrix}.
The binary value signifies whether that third-party exists in at least one topical subpage of the given topic or not.

As we also want to understand the semantic relation between topical subpages across websites, we applied content analysis (\textit{Step 9}).
First, we crawled the HTML of the 1387 URLs, obtaining the HTML source for 1349.
Then, we extracted the textual content from each and applied an NLP Pre-processing pipeline (case-folding, eliminating blank spaces, word segmentation, un-accenting words, removing punctuation, lemmatization and stop-word removal).
Then, we computed a \texttt{tf-idf} vector for each topic, by combining all extracted text from each subpage into one document per topic, thus, constructing a 16x25168 matrix with 16 topics and 25168 features.
We refer to this as \textit{Content Matrix}.
We use both aforementioned matrices in Section~\ref{sec:results-clustering}.

\section{Analysis of Tracking on Topical Subpages}
\label{sec:topical_tracking}

In this section, we analyse and compare the tracking found on Indian news homepages and their topical subpages, in order to find differences between them (Sec.~\ref{sec:results-difference-tracking}), types of third-parties involved (Sec.~\ref{sec:results-tracking-categories}), dominant tracking entities (Sec.~\ref{sec:results-top-thirdparties}), preference of them to track specific topical subpages (Sec.~\ref{sec:results-preference-tracking}) and how this compares with content (Sec.~\ref{sec:results-clustering}).
Where appropriate, we use the Kolmorogov-Smirnov (KS) test to assess statistical significance of distribution differences.

\subsection{\textbf{Is tracking on topical subpages different from homepage?}}
\label{sec:results-difference-tracking}
\begin{figure*}
    \begin{subfigure}{0.45\textwidth}
    \centering
      \includegraphics[width=\linewidth]{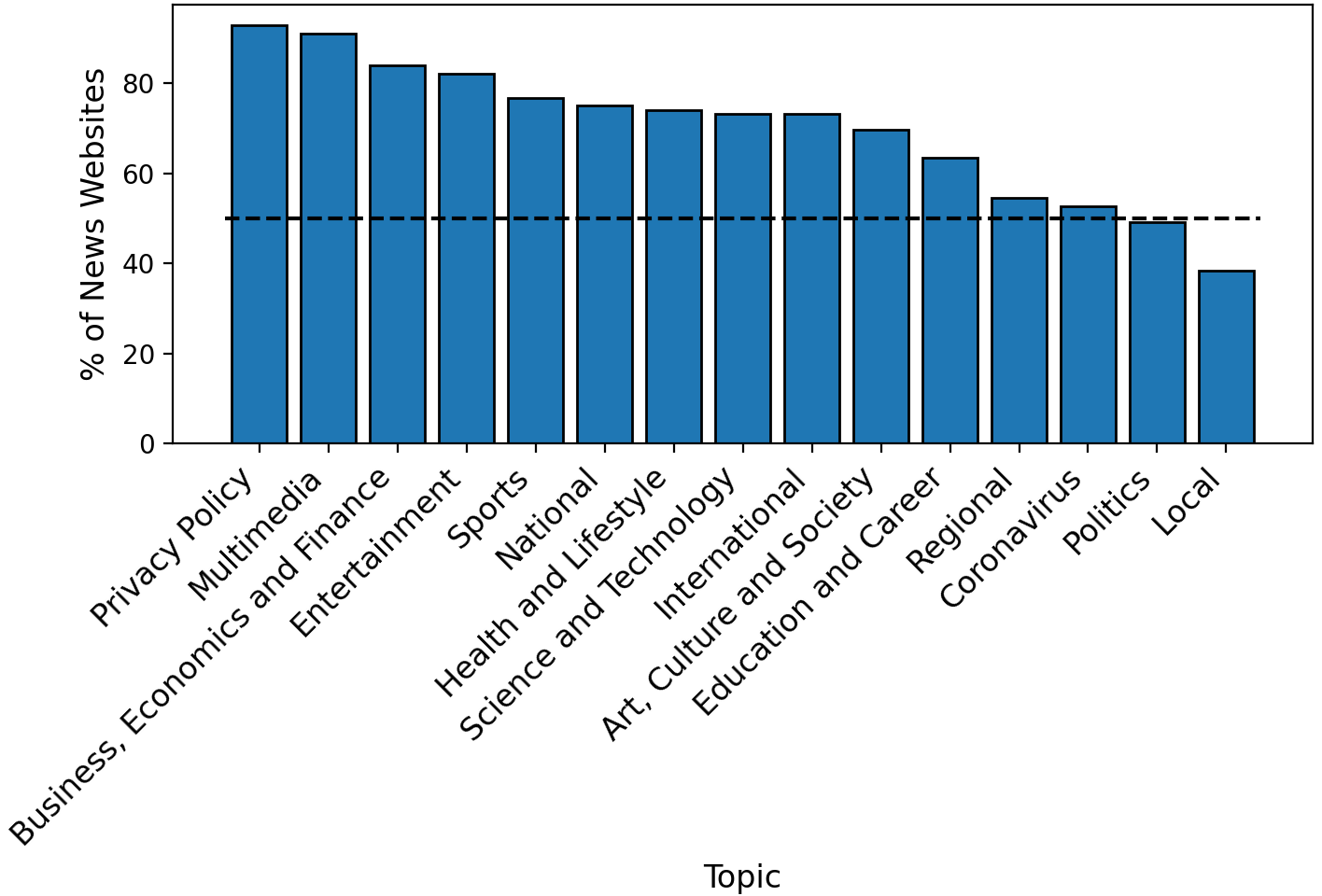}
      %\caption{Percentage of Websites covered by each topic.}
    \end{subfigure}
    \begin{subfigure}{0.45\textwidth}
    \centering
      \includegraphics[width=0.92\linewidth]{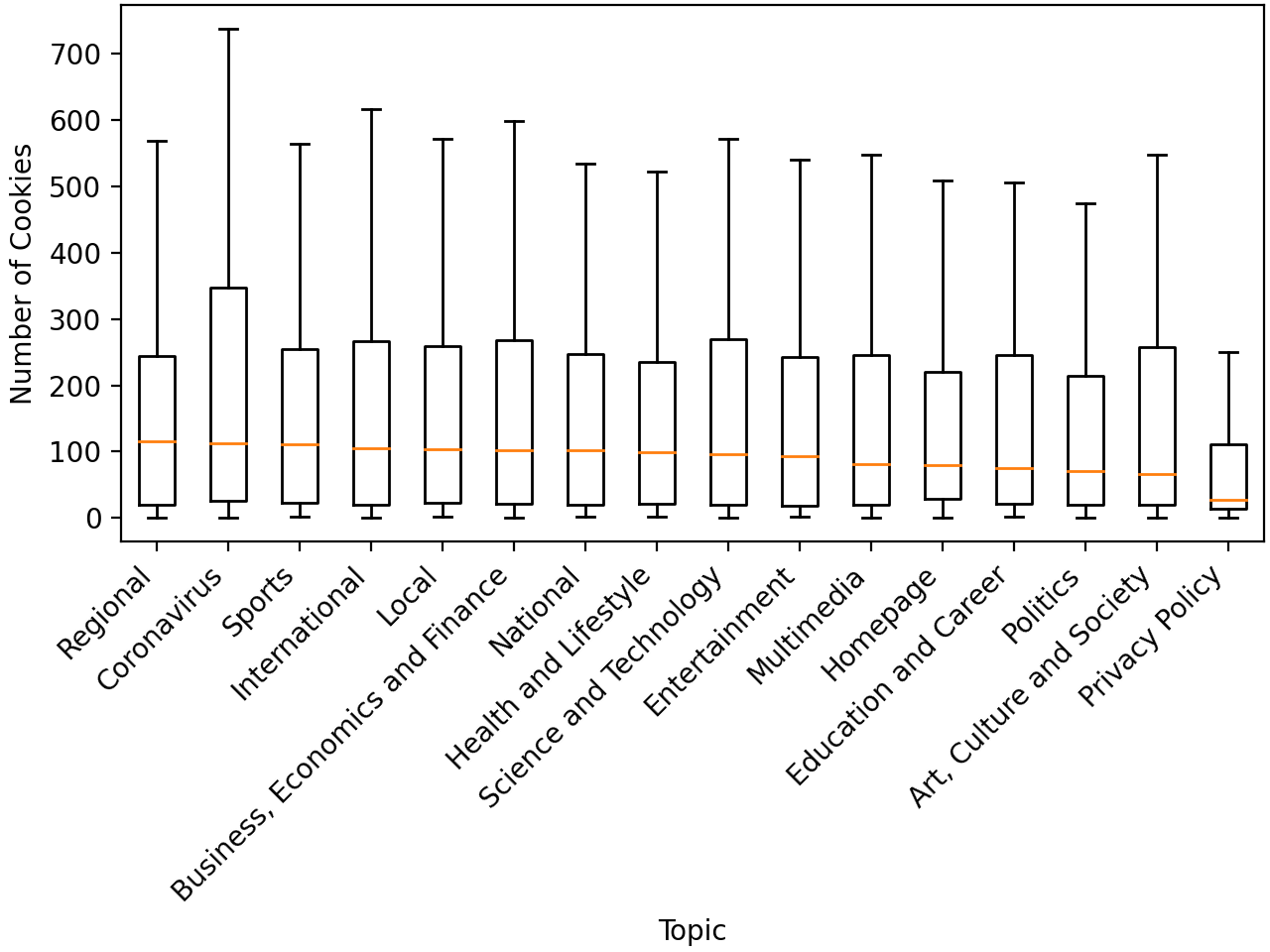}
      %\caption{Box plot of number of cookies for each topic.}
    \end{subfigure}
    \caption{Left: Portion of websites covered by each topic. Right: Number of cookies set in topical subpages across websites.}
    \label{fig:topic_percent_sites_fig:topic_median_cookies_boxplot}
\end{figure*}

To understand the difference between homepage and topical subpages, we first present the existence of topical subpages across news websites in our datset. 
Second, we use the number of cookies injected on these webpages as a proxy for tracking as previous studies~\cite{englehardt2016census,agarwal2021under}.
Figure~\ref{fig:topic_percent_sites_fig:topic_median_cookies_boxplot} (Left) shows the percentage of Indian news websites covered by each topic.
Clearly, not all topics are covered by all news websites.
\textit{Privacy Policy} appears in 92\% of websites, perhaps as an attempt to comply with EU/USA laws.
\textit{Multimedia} is in 86\% of websites, as users may be interested in news with media content like videos.
We also see `hot' topical pages like \textit{Coronavirus}, mainstream topics like \textit{Politics}, and city-related topic like \textit{Local},which are covered by nearly one in every two news websites.

Figure~\ref{fig:topic_percent_sites_fig:topic_median_cookies_boxplot}~(Right) shows the box plot of number of cookies for each topical subpage, ordered by their median number of cookies.
We also include their homepage for comparison.
We find that 11 out of 15 topics have higher median cookies than the Homepage, indicating more intense tracking on these topical subpages.
Interestingly, less frequent topics (occurring in half of websites) such as \textit{Regional} and \textit{Coronavirus} have the highest and second highest number of median cookies, respectively.
Also, \textit{Coronavirus} has the highest mean number of cookies (243), which can have up to 1280 maximum cookies for a single webpage on this topic.
\textit{Privacy Policy} has the least median number of cookies (28). Interestingly, eight websites do not have privacy policy pages.
The remaining 104 websites have one but in English only\footnote{We detect the language of topical pages using `Langdetect' which is built upon Google's language-detection  \url{https://pypi.org/project/langdetect/}.}.
In fact, in our dataset, 23 out of 104 websites are in Hindi or other regional languages, but they all have privacy policy pages in English.
This suggests that privacy policy subpages in news websites are mostly acting as placeholders and do not attract many users or trackers~\cite{krumay2020readability}.

\noindent\textbf{Findings:} Nearly 75\% of topical subpages track users more intensely than Homepages.
In some cases, this tracking using cookies can be at up to 1.4x than Homepage.
Topics \textit{Regional} and \textit{Coronavirus} have the highest and second highest number of cookies, respectively, and have statistically significant more tracking than Homepages with KS-value 0.33 (p-value: 0.05) and 0.18 (p-value: 0.07), respectively.
\textit{Privacy Policy} topic tracks users the least with 65\% fewer cookies than Homepages.
Also, eight websites do not have privacy policy page and wherever it exists, it is in English. 

\subsection{What are the categories of third-party cookies tracking different topical subpages?}
\label{sec:results-tracking-categories}

\begin{figure*}
\centering
    \begin{subfigure}{0.5\textwidth}
     \centering
      \includegraphics[width=\linewidth]{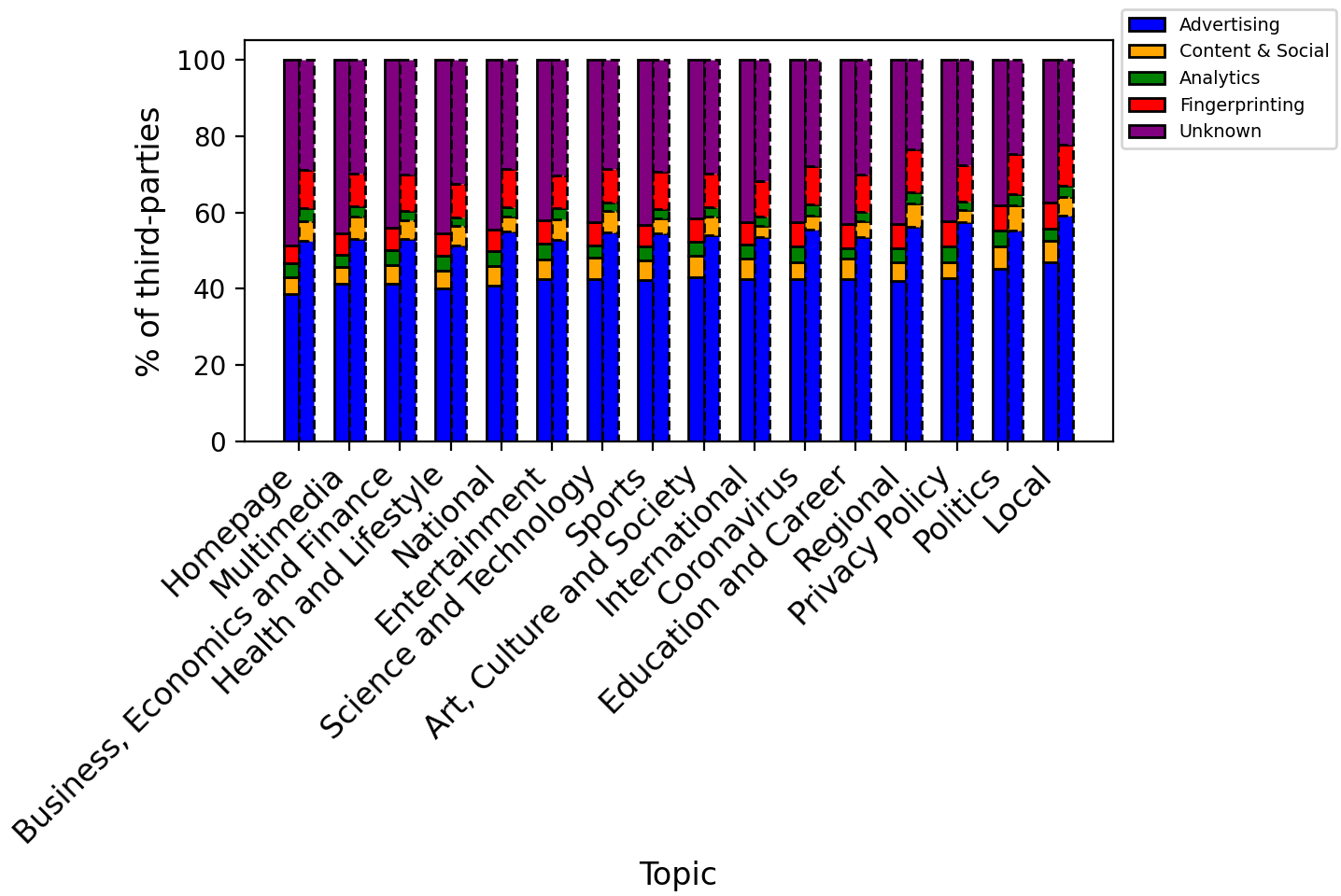}
      \end{subfigure}
    \begin{subfigure}{0.45\textwidth}
    \centering
      \includegraphics[width=\linewidth]{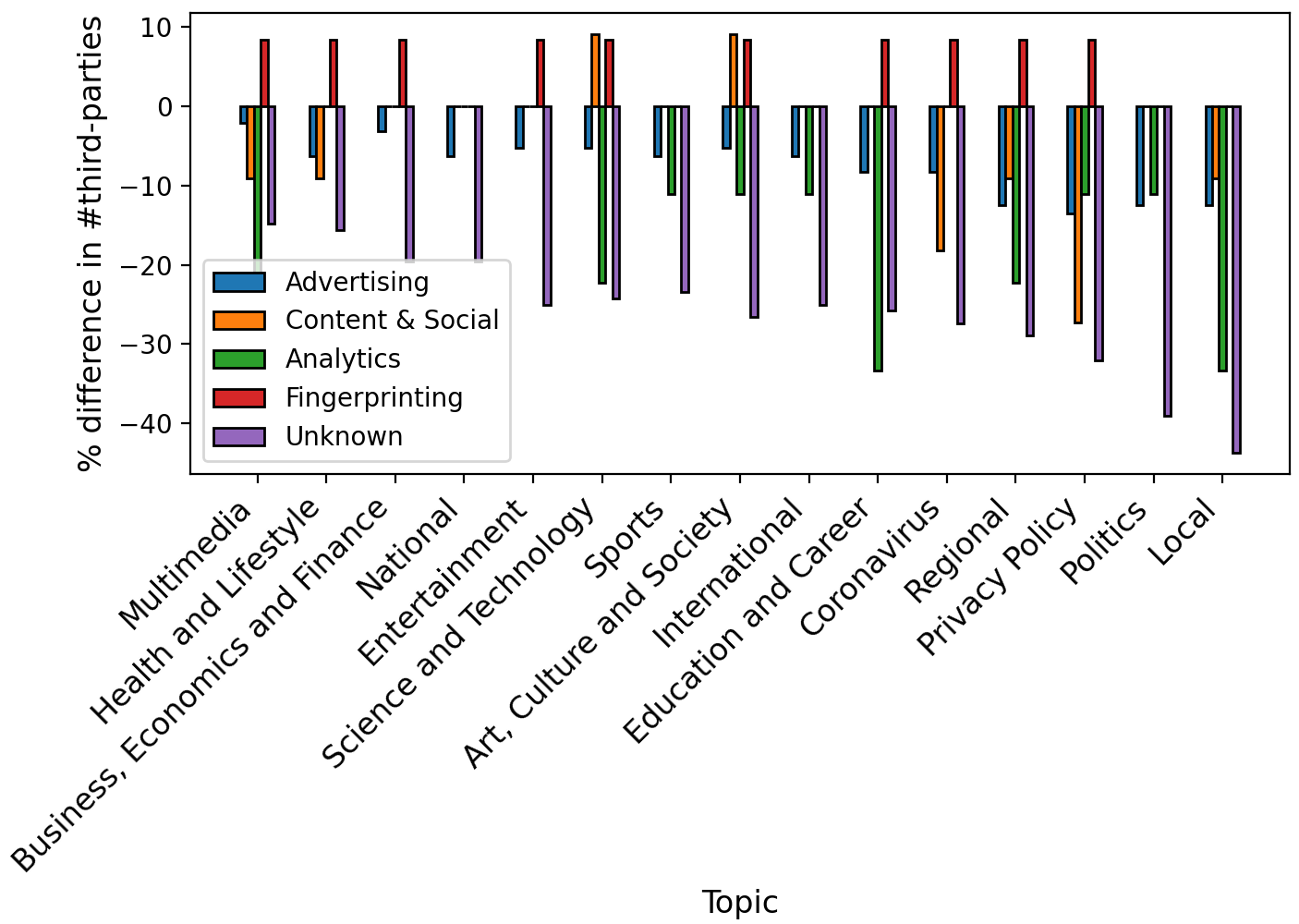}
    \end{subfigure}
    \caption{Left: Topic-wise cookie categorization using Disconnect List for all news websites (solid bars) and top 10 news websites (dashed bars). Right: Topic-wise percentage difference in number of third-parties for each category w.r.t. Homepage.
      }
    \label{fig:cookie_categories_fig:topic_per_diff_cookies}
\end{figure*}

Next, we study the type of cookies set additionally to Homepage.
We breakdown the cookies into types using domains list by \emph{Disconnect.me} (DL)~\cite{disconnect-list}.
This list provides categories of third-party (TPs) like \emph{Advertising, Content \& Social, Analytics} and \emph{Fingerprinting}.
Figure~\ref{fig:cookie_categories_fig:topic_per_diff_cookies} (Left) shows the bar plot of topic-wise cookie categorization for all news websites (represented via solid bars) and top ten popular (top as per Facebook followers) websites (represented via dashed bars).
Topics are sorted in descending order of number of distinct third-parties in all webpages.
Note that \textit{Unknown} represents the number of TPs that cannot be categorized using DL.
More than 80\% of TP domains categorized by DL belong to \textit{Advertising}.
This is expected since most news websites generate income through display ads.
Interestingly, the second largest portion of TPs belongs to \textit{Fingerprinting}.
In the case of all websites (solid bars), \textit{Homepage} has the highest number of distinct TPs followed by other topics.
But in the case of top ten websites, it is not so: nine out of 16 topics have more distinct TPs than Homepage, with \textit{Health and Lifestyle} having the highest number with KS-value 0.8 (p-value: 0.04).
This shows that, in general, more tracking is performed by more distinct TPs on the topical subpages.
Surprisingly, we find that 182 \textit{Advertising} and other TP domains are tracking users on \textit{Privacy Policy} pages.
The \textit{Unknown} TPs are mostly India-specific, but few belong to other countries as well: e.g., \textit{audiencemanager.de} (Germany), \textit{gsspat.jp} and \textit{fw-ad.jp} (Japan) etc. Majority of the TPs are the ones that have preferential attachment towards some topics like \textit{asianpaints.com} for Coronavirus, \textit{gamepix.com} for Sports etc. Interestingly, two TP domain names (\textit{s3xified.com} and \textit{xvidoes.com}) appear to be related to Adult websites. We also notice that for top ten news websites, the amount of  \textit{Unknown} is less when compared to all news websites.

Figure~\ref{fig:cookie_categories_fig:topic_per_diff_cookies} (Right) shows the bar plot of percentage difference between Homepage and subpages of each type of third-party.
We find that ten out of 15 topics, such as \textit{Multimedia}, \textit{Health \& Lifestyle}, and \textit{Coronavirus}, have more \textit{Fingerprinting} TP domains than the Homepage.
Moreover, \textit{Science \& Technology} and \textit{Art, Culture \& Society} also have more \textit{Content \& Social} TPs than the Homepage.

\noindent \textbf{Findings:} More than 80\% of cookies set on news websites, as categorized by DL, are \textit{Advertising}.
The fraction of \textit{Unknown} cookies is less in top ten popular websites, as they probably employ more well known and categorized TP trackers.
\textit{Advertising} and other TPs track users on \textit{Privacy Policy} pages as well.
In case of top ten news websites, the majority of topics have more distinct TPs than Homepage, with \textit{Health \& Lifestyle} having the highest number with KS-value 0.8 (p-value: 0.04).
In general, topics have higher \textit{Advertising} and \textit{Fingerprinting} cookies than Homepages with KS-value 0.9 (p-value: 0.05) and 0.67 (p-value: 0.04) respectively.
Among the TPs with \textit{Unknown} category, few domain names are related to Adult-content websites, while some TPs belong to other countries like Japan, Germany etc.

\subsection{Are top third-party trackers present in all the places?}
\label{sec:results-top-thirdparties}

\begin{figure*}
    \begin{subfigure}{0.5\textwidth}
    \centering
      \includegraphics[width=\linewidth]{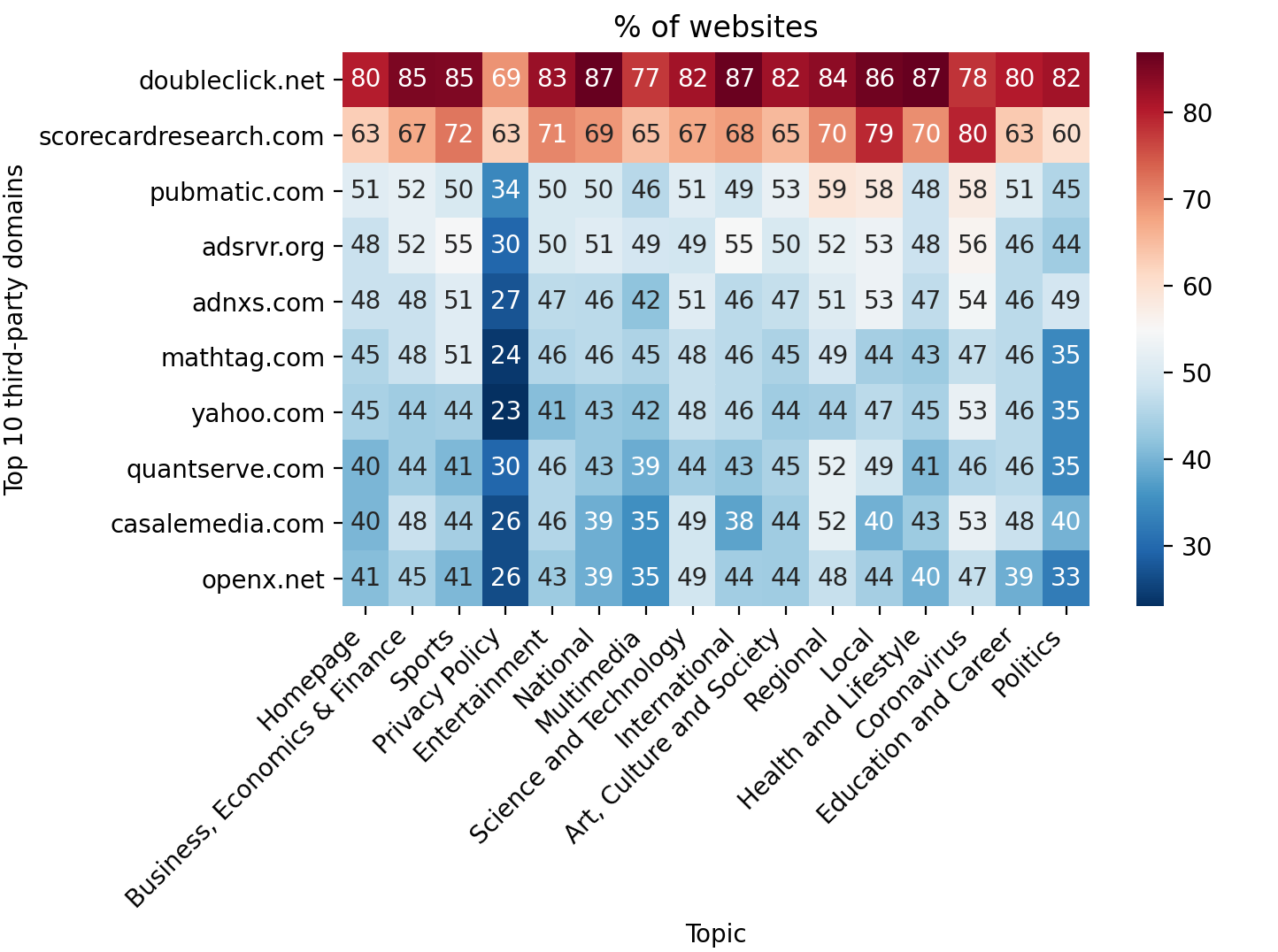}
    \end{subfigure}
    \begin{subfigure}{0.45\textwidth}
    \centering
      \includegraphics[width=\linewidth]{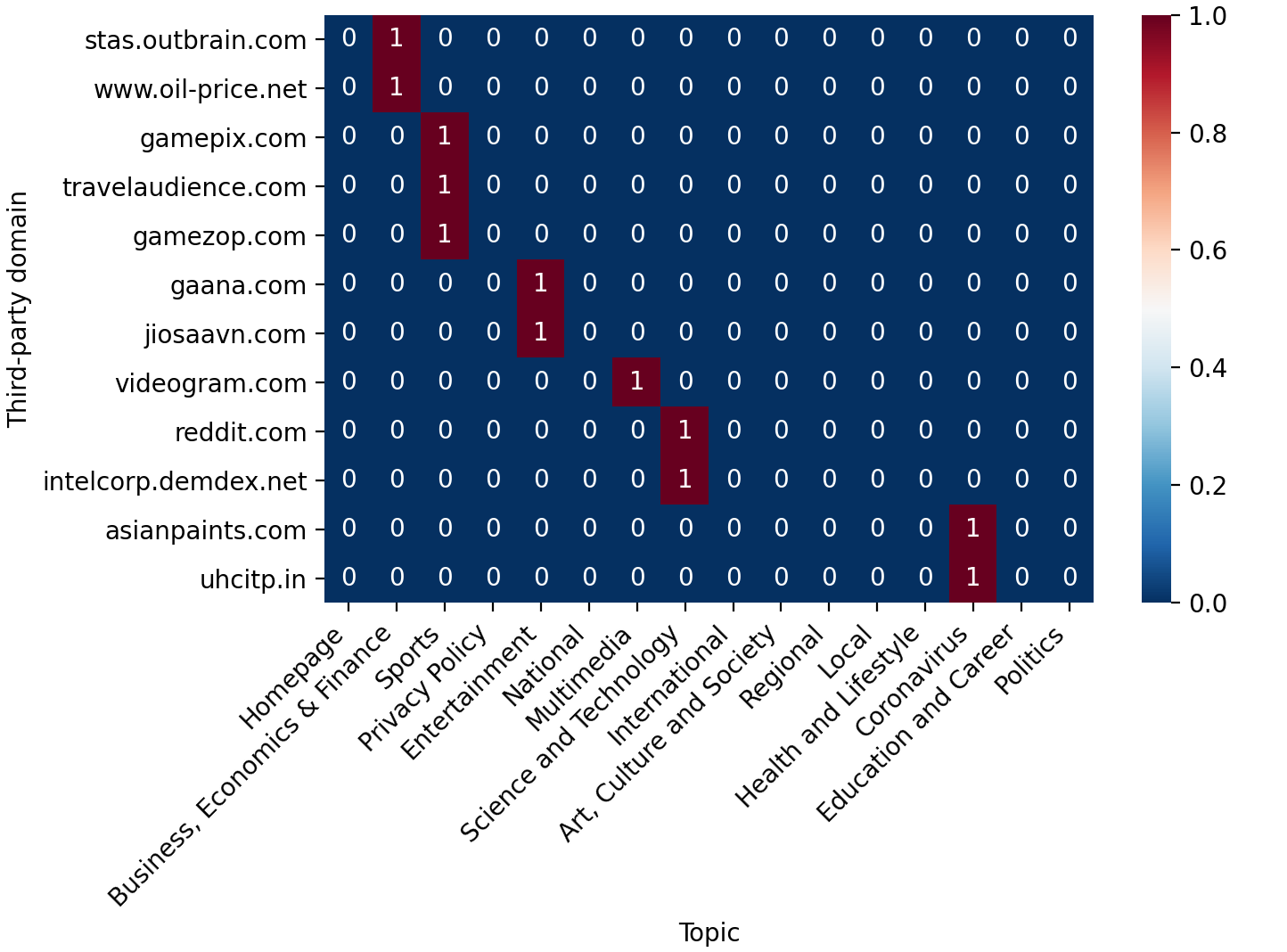}
    \end{subfigure}
    \caption{Left: Heat map of top 10 third-parties. Right:
    Heat map of top 12 preferentially attached third-parties to topics.
    }
    \label{fig:tp_topicwise_heatmap_fig:tp_exceptions_heatmap}
\end{figure*}

In total, 523 distinct TP domains are found setting cookies on news websites and their topical subpages.
The top 20 TPs are almost common in all topics.
Due to space, we focus on the top ten TPs.
In Figure~\ref{fig:tp_topicwise_heatmap_fig:tp_exceptions_heatmap} (Left) shows the heat map of percentage of topic subpages in which the top ten TPs appear.
Like past studies that have found a handful of top TP trackers present in most websites~\cite{englehardt2016census,agarwal2021under,agarwal2020stop}, we also find that the top ten TP trackers are in large fraction of the topical subpages and Homepages in our dataset.
Interestingly, all pages are found to have some of the top ten TPs with little difference in the TP ranking (based on the number of cookies set).
The top TP \textit{doubleclick.net} is present in more than 75\% of webpages for all topics except \textit{Privacy Policy}.
From rank 3 onward (i.e., \textit{pubmatic.com}), we see a sudden drop in and then fairly consistent amount of websites coverage (40\% to 50\%).
As reported in Sec~\ref{sec:results-difference-tracking}, \textit{Privacy Policy} has the lowest number of median cookies, and even the top ten TPs are least interested, as they cover the lowest portion of \textit{Privacy Policy} pages.

\noindent \textbf{Findings:} A handful of top third-parties track users not only on Homepages, but follow them on specific topical subpages as well.
Top 20 TPs are almost common in all topics studied.
The top TP \textit{doubleclick.net} is present in more than 75\% of webpages for all topics except \textit{Privacy Policy}.

\subsection{Are third-party trackers preferentially attached to specific topical subpages?}
\label{sec:results-preference-tracking}

As we saw earlier, 11 out of 15 topics have higher median cookies than the Homepage, but TPs are not evenly present in all topics.
Thus, here we study {\it which TPs} track {\it what} specific topical subpages and if there is preferential attachment of TPs towards topics.
Figure~\ref{fig:tp_topicwise_heatmap_fig:tp_exceptions_heatmap} (Right) shows the heat map of 12 TPs preferentially attached (100\% presence) to specific topics.
The `1' in a cell represents that a TP is present only in that specific topic.
Overall, we find 12\% (64 in 523) TPs are preferentially attached to just one of the 15 topics.
Few interesting examples are described below:
\vspace{-0.2cm}
\begin{itemize}
    \item Entertainment: \textit{gaana.com} and \textit{jiosaavn.com} are music websites/TPs and, with 3 other TPs, are present only here.
    \item Sports: \textit{gamepix.com} and \textit{gamezop.com}, with 6 others, are present only in this topic.
    \item Coronavirus: \textit{asianpaints.com}, which recently released an \emph{Anti-COVID-19} paint, is found only in this topic. Also, \textit{uhcitp.in}, a Universal Health Coverage website, is tracking Coronavirus topic only. 10 other TPs are also present.
    \item Business, Economics, \& Finance: \textit{oil-price.net}, a website for oil prices, with 4 others, is present only in this topic.
    \item Multimedia: \textit{videogram.com}, a video analytics website boosting video revenue, with 4 others, is present here.
    \item Science \& Technology: \textit{intelcorp.demdex.net}, a technology company, is specifically tracking this topic. \textit{reddit.com} is also present with 2 other TPs.
\end{itemize}

\noindent \textbf{Findings:} Specialized third-parties appear to track specific subpages.
In fact, there is (100\%) preferential attachment of 12\% third-parties towards specific topics.
This is due to third-parties tracking users' interest on specific news topics, thus, trying to optimize their own marketing and ad-targeting campaigns, or selling user profiling to other TPs not present.

\subsection{Unsupervised Clustering of Topics: Do topical subpages show relatedness based on the hosted content or third-party trackers?}
\label{sec:results-clustering}

While exploring topical tracking, it becomes important to study the two complementary aspects of this ecosystem:

\begin{enumerate}
    \item[Users:] The actual content hosted by different topical subpages, and consumed by users.
    Our aim is to understand if there are semantically similar topical groupings among the subpages.
    For this, we analyze the 16 x 25168 tf-idf content matrix (Sec.~\ref{sec:methodology-crawling}).
    \item[TPs:] The third-parties are embedded in these subpages, and are silently tracking these users.
    We try to understand if TPs are tracking users of specific profiles, i.e., users that are interested in specific combinations of topics.
    For this, we analyze the 16 x 651 TP tracking matrix (Sec.~\ref{sec:methodology-crawling}).
\end{enumerate}

\begin{figure*}
  \begin{subfigure}{0.5\textwidth}
    \centering
      \includegraphics[width=\columnwidth]{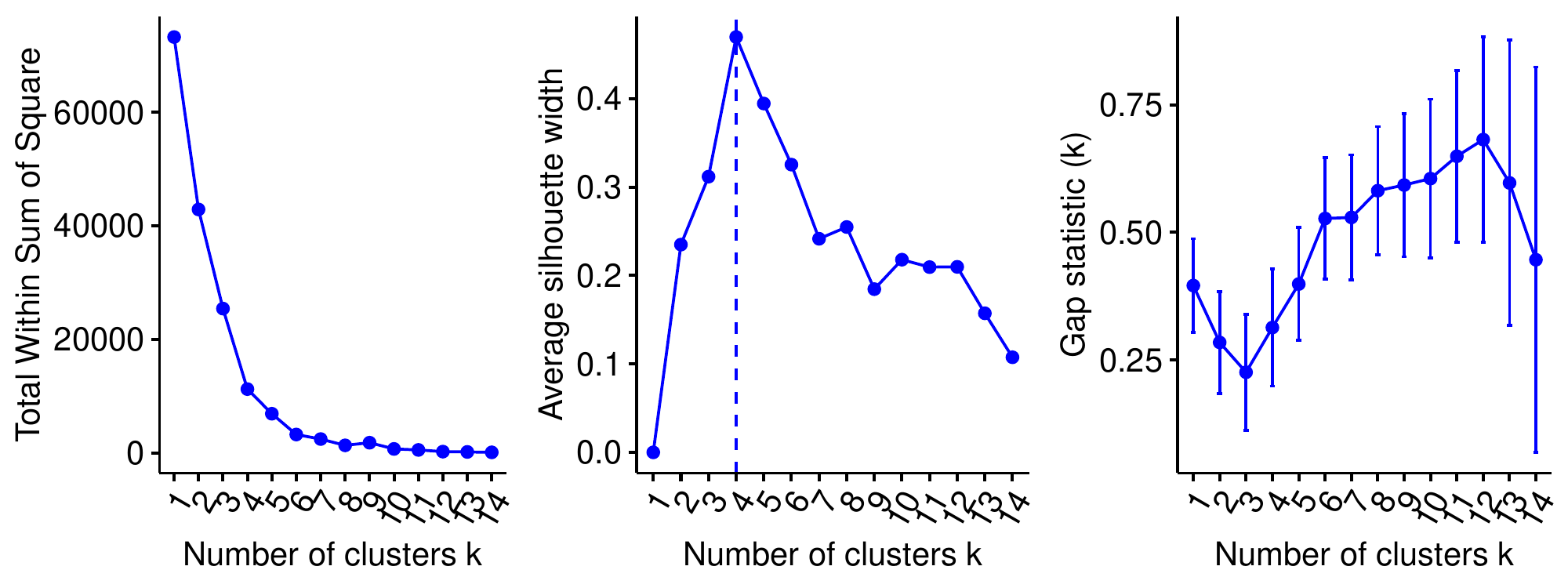}
      \caption{Clustering metrics for K-means on PCA-reduced $N=2$ and different values of K on content matrix.}
      \label{fig:content_clustering_metrics}
    \end{subfigure}
    \begin{subfigure}{0.45\textwidth}
    \centering
      \includegraphics[width=\columnwidth]{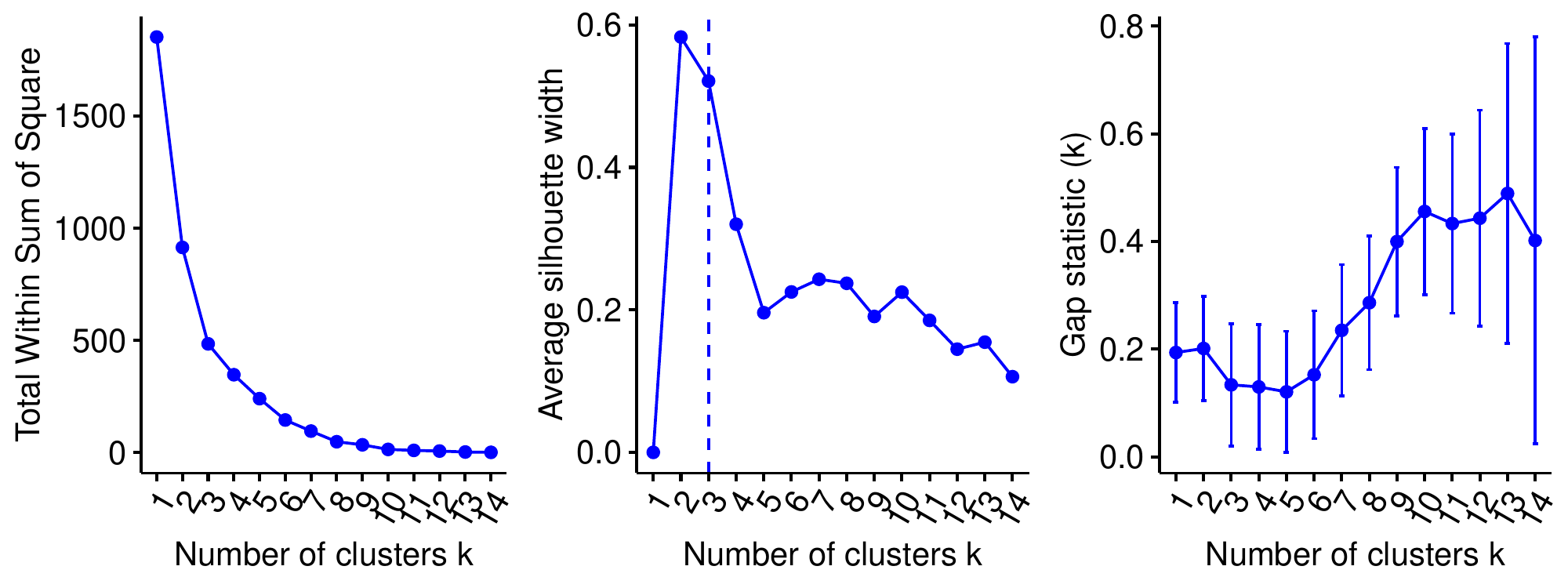}
      \caption{Clustering metrics for K-means on PCA-reduced $N=2$ and different values of K on tracking matrix.}
     \label{fig:tracking_clustering_metrics}
    \end{subfigure}
    \begin{subfigure}{0.45\textwidth}
    \centering
     \includegraphics[width=\columnwidth]{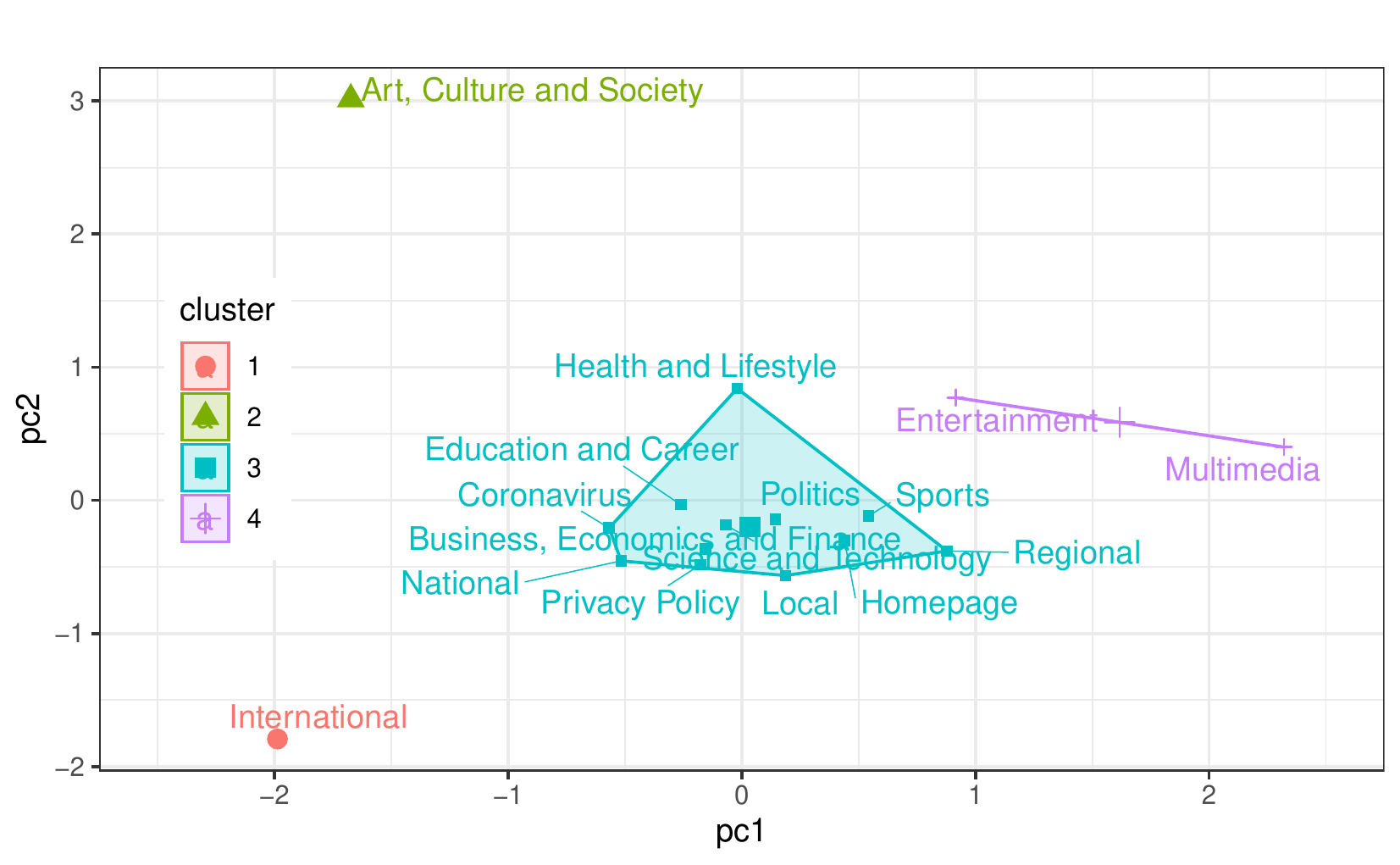}
      \caption{K-Means clustering on tf-idf-based content features PCA-reduced to 2 dimensions and $K=4$.}
      \label{fig:content_clustering}
    \end{subfigure}
    \begin{subfigure}{0.45\textwidth}
    \centering
     \includegraphics[width=\columnwidth]{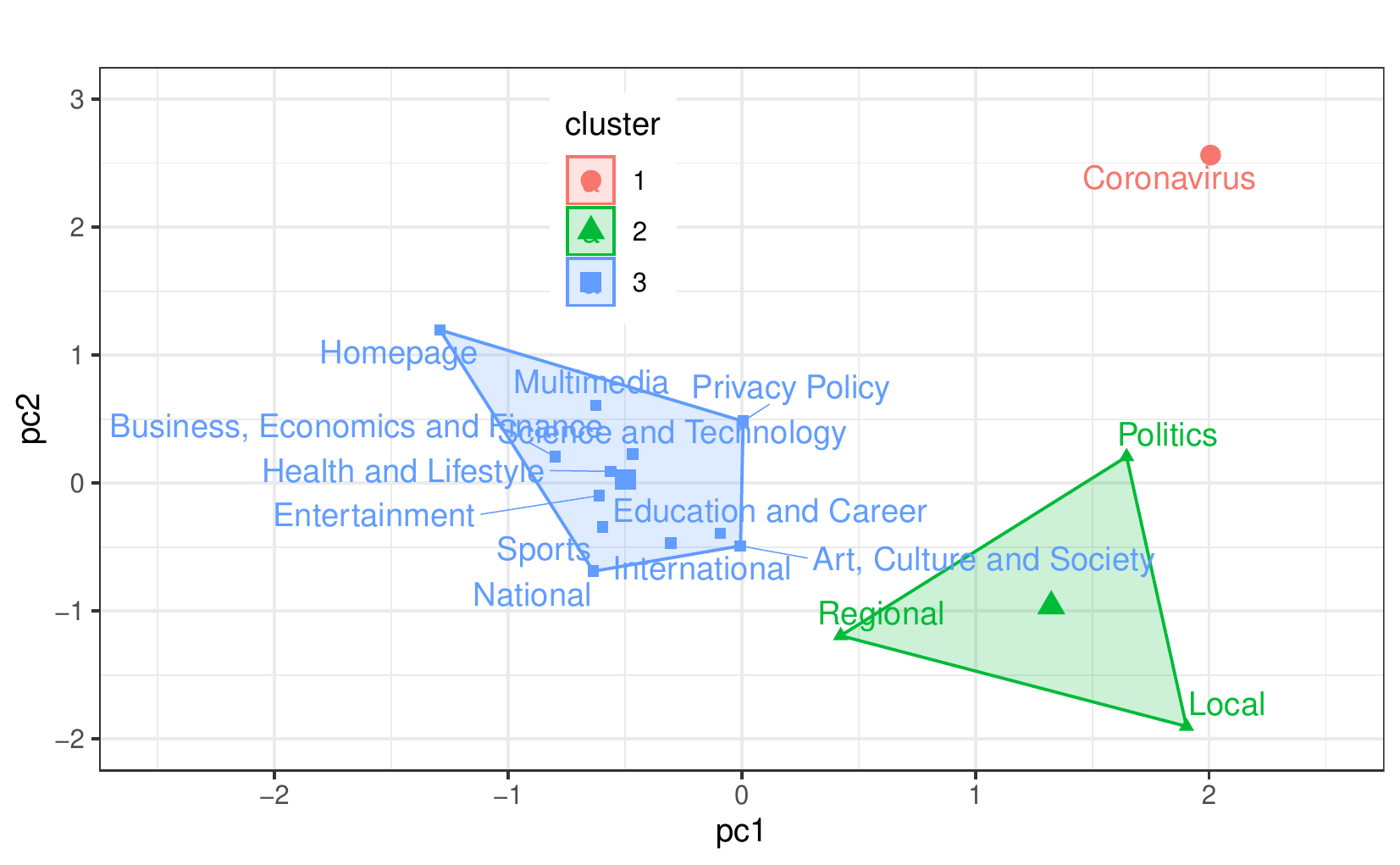}
      \caption{K-Means clustering on third-party-based tracking features PCA-reduced to 2 dimensions and $K=3$.}
      \label{fig:tracking_clustering}
    \end{subfigure}
    \caption{Top: Clustering Quality Metrics: SSE (elbow method), Silhouette Coefficient, and Gap Statistic, for K-means on PCA-reduced data for shown values of N and K in (a) Content-based tf-idf features and (b) Tracking-based third-party features.
    Bottom: Visualization of Clustering on (c) Content-based tf-idf features and (d) Tracking-based third-party features.}
\end{figure*}

We explore such patterns by performing unsupervised K-means clustering on the content and tracking matrices.
To reduce noise due to high feature dimensionality, we map the features of both matrices to a lower dimension ``N'' using PCA.
We observed that PCA-reduced $N=15$ features cover 100\% variance of the original data in both matrices.
Thus, we perform K-means clustering for different combinations of (N, K), with $N, K \in [2, 15]$.
To guide our selection for an appropriate N and K from all combinations, we compute three standard clustering quality metrics~\cite{yuan2019research,subramani2020measuring}: SSE (elbow curve), silhouette coefficient, and gap statistic.
Figures~\ref{fig:content_clustering_metrics} and~\ref{fig:tracking_clustering_metrics} show these metrics across K, for the most appropriate value of $N=2$, for both content- and tracking-based clustering, respectively.
As evident, for content clustering (Fig.~\ref{fig:content_clustering_metrics}), we obtain the highest silhouette score for $N=2$ and $K=4$. Silhouette score indicates the ``tightness'' of clusters~\cite{subramani2020measuring}.
Additionally, elbow also occurs at $K=4$.
Similarly, among all combinations of N and K, the highest silhouette coefficient for tracking-based clustering (Fig.~\ref{fig:tracking_clustering_metrics}) occurs at $N=2$ and $K=2$. 
However, $K=2$ has a higher SSE and hence we choose $N=2$ and $K=3$ for clustering. 
$K=3$ is also optimal as per the elbow curve.
Overall, $N=2$ preserves $\approx30\%$ variance for both content and tracking matrices.

Figure~\ref{fig:content_clustering} shows the resultant clusters obtained by performing K-means with $K=4$ on content.
Many of the topics are clustered together, revealing a possible similar use of language across the board.
Content in \textit{Entertainment} and \textit{Multimedia} subpages, although different from other topics, is very similar to each other.
Finally, \textit{Art, Culture, and Society}, as well as \textit{International} seem to be prominently different from the content of other topics, and are assigned to their own cluster.
Figure~\ref{fig:tracking_clustering} depicts the other side: clusters based on TP tracking.
We observe that \textit{Local}, \textit{Regional} and \textit{Politics} have been tracked differently to other topics, but similarly with each other.
This means same trackers are interested in users viewing these topics together.
This can be due to the local and regional political elections happening in different parts of India during the time of crawling, which could have attracted similar trackers.
Interestingly, \textit{Coronavirus}, which is a transient trend, is tracked most differently from the conventional topics.
This is probably because there are a lot of new and unexpectedly higher number of trackers investing on topical subpages of \textit{Coronavirus} than others, since it has become a hot topic.
Also, Homepage, although it is present in cluster 1, is expected to have a wide variety of trackers embedded and thus, it is clearly far from other topics.

\noindent \textbf{Findings:}
Overall, the two sides of clustering revealed different views.
Users, due to content they are presented, can perceive the majority (75\%) of topical subpages in the same way.
Many of them have similar textual content and are clustered together, with the exceptions of \textit{Art, Culture, and Society}, \textit{International}, \textit{Entertainment} and \textit{Multimedia}, with the last two being clustered together due to very similar content.
On the contrary, embedded TP trackers, tend to track specific subpages simultaneously, revealing possible user profiling in action (for e.g., users interested in \textit{Local}, \textit{Regional} and \textit{Politics}).

\section{Conclusion and Future Work}

Our study is the first step in trying to understand the ``Wild East'' of web tracking in the Indian context. We aimed to cover two gaps in the tracking ecosystem.
First, operationalizing the identification of topics using URLs of topical subpages.
Second, understanding tracking on the identified topical subpages of the Indian news websites.
In general, we find that the characteristics of tracking are different on topical subpages when compared to Homepages of the same news websites.
We summarize our findings as follows:
\begin{itemize}
    \item Our proposed DiBETS model performs well for identification of news topics using only the URLs as input, with minimal user interventions. 
    \item The tracking on topical subpages of Indian news websites are substantially different from their Homepage: more than 66\% of topics have more median number of cookies than Homepages. 
    \item In case of top ten news websites, the majority of topics have more distinct third-parties than Homepage, and in general, topics have higher Advertising and Fingerprinting cookies than Homepages.
    \item We found third-parties that appear to track specific subpages. There are 12\% of third-parties having 100\% preferential attachment towards specific topics. Further, there is topical relatedness when it comes to third-party trackers' interest in users who view a set of topics together.
    \item Privacy policy subpages are mostly acting as placeholders with minimal interest of trackers. We found that eight websites do not have privacy page (also confirmed manually). All other websites have privacy policy in English, even the ones in regional languages.
    \item Large proportion of trackers are from the category of advertising. For some topics, there are additional trackers, which are not on Homepage. This mostly consists of trackers which indulge in user fingerprinting.
    \item Google's \textit{doubleclick.net} covers nearly 75\% of the news websites in our data.
    \item Content shown to users appears similar for topical subpages. However, embedded third-party trackers tend to track specific subpages simultaneously, revealing possible user profiling in action (for e.g., users interested in \textit{Local}, \textit{Regional} and \textit{Politics}), or new topics such as \textit{Coronavirus}.
\end{itemize}

\noindent \textbf{Ethics and Limitations:} Our study does not involve any personal data from any real user.
Moreover, the study helps in raising awareness and the need for appropriate laws and restrictions for tracking in the `East'.
We plan to release all our data, code and methods for reproducibility and extensibility purposes. 
We tried to make our study as complete as possible. In doing so, there exist some limitations as well.
The topic classification is probability-based and sensitive to threshold setting.
The analysis is based on one single (best matched) subpage per topic per website.
DiBETS users may be required to manually label keywords for adding them to topical dictionary.
The findings may be sensitive to the choice of day, time and place of crawling due to the dynamic ecosystem of tracking.

\noindent \textbf{Future Work:} We aim to expand our news websites list. 
Using our DiBETS model, we aim to understand topical subpage existence along party lines (Right versus Left) of news. 
Prior studies have shown that news leanings can be attached to the content they publish for users~\cite{entman2007framing,haselmayer2017partisan}. We also aim to understand the prioritisation of topical news sharing on social media in the Indian context~\cite{agarwal2020characterising}.

\begin{acks}
\noindent N. Kourtellis has been partially supported by the European Union’s Horizon 2020 Research and Innovation Programme under grant agreements No 830927 (Concordia), No 871793 (Accordion), and No 871370 (Pimcity).
These results reflect only the authors' findings and do not represent the views of their institutes/organisations.
\end{acks}

\bibliographystyle{ACM-Reference-Format}
\bibliography{sample-base}

\end{document}